\documentclass[twocolumn,prb,superscriptaddress,showpacs,amsmath,amssymb]{revtex4}
\usepackage{graphicx}
\usepackage{dcolumn}
\usepackage{bm}

\begin{document}
\title{Polaron Transport in the Paramagnetic Phase of Electron-Doped
Manganites}
\author{J.~L.~Cohn}
\affiliation{Department of Physics, University of Miami, Coral
Gables, Florida 33124}
\author{C.~Chiorescu}
\affiliation{Department of Physics, University of Miami, Coral
Gables, Florida 33124}
\author{J.~J.~Neumeier}
\affiliation{Department of Physics, Montana State University,
Bozeman, Montana 59717}

\begin{abstract}
The electrical resistivity, Hall coefficient, and thermopower as
functions of temperature are reported for lightly electron-doped
Ca$_{1-x}$La$_x$MnO$_3$($0\leq x\leq 0.10$). Unlike the case of
hole-doped ferromagnetic manganites, the magnitude and temperature dependence of
the Hall mobility ($\mu_H$) for these compounds is found to be
inconsistent with small-polaron theory. The transport data are
better described by the Feynman polaron theory and imply intermediate
coupling ($\alpha\simeq 5.4$) with a band
effective mass, $m^{\ast}\sim 4.3 m_0$, and a polaron mass,
$m_p\sim 10m_0$.

\end{abstract}

\pacs{PACS numbers: 75.47.Lx, 71.38.-k, 72.20.-i}
\maketitle

\section{\label{sec:Intro} INTRODUCTION}

Studies of perovskite manganites have drawn considerable attention
to the issue of electronic phase separation as a paradigm for
understanding strongly correlated electron systems generally.\cite{Dagotto}
Inhomogeneous magnetic ground states, observed over a broad range
of compositions, are thought to arise from inhomogeneous paramagnetic
states as a consequence of competing interactions. In the case of
hole-doped, ferromagnetic colossal magnetoresistance (CMR) manganites,
there exists compelling evidence that the paramagnetic phase is small-polaronic,\cite{Jaime,CohnReview,SalamonJaime}
and inhomogeneous due to local ferromagnetic (FM) and charge-order fluctuations.

Lightly electron-doped manganites, such as
Ca$_{1-x}$La$_x$MnO$_3$, are structurally simpler than
their hole-doped counterparts since cooperative Jahn-Teller distortions
are largely absent due to the tetravalent state of most Mn ions.
Antiferromagnetic (AF) superexchange interactions are dominant
and dictate a G-type AF ground state for CaMnO$_3$ below $T_N=125\ K$.
Electron doping\cite{Maignan,NeumeierCohn,OtherRaveau,YCa} via substitution of trivalent ions for Ca
induces a weak ferromagnetic (FM) moment associated with an inhomogeneous magnetic
state.\cite{LingGranado}

Transport properties in the paramagnetic phase of electron-doped manganites differ substantially from
those of hole-doped FM compositions and consensus about the conduction mechanism is lacking.
Both the resistivity and thermopower are thermally activated
in hole-doped compounds in a manner consistent with a thermally activated mobility and
small polaron theory.
In contrast, the resistivity of lightly electron-doped compounds exhibits a positive temperature coefficient
at $T\geq$~150-200~K, and the thermopower decreases with decreasing $T$.\cite{Maignan,NeumeierCohn,OtherRaveau,YCa}
The binding energy of small
polarons, inferred from transport, decreases with increasing hole concentration\cite{CohnReview,PolaronEnergy},
suggesting the possibility of a qualitative change in the character of polarons for electron-doped compositions.

The present paper focuses on charge carrier transport (resistivity, Hall effect, thermopower) near $T_N$
and above in Ca$_{1-x}$La$_x$MnO$_3$ with the goal of clarifying the role and nature of polarons in the
paramagnetic phase of electron-doped materials.  The Hall mobility, which has
not been previously reported for such compositions,
is found to favor a large-(continuum) rather than a small-polaron picture.
Analysis of both the mobility and thermopower imply intermediate coupling, with
electron-phonon coupling parameter, $\alpha\simeq 5.4$, an electron effective mass, $m{\ast}\sim 4.3m_0$,
and a polaron mass, $m_p\simeq 10m_0$.
%

\section{\label{sec:Expt} EXPERIMENT}

Polycrystalline specimens of Ca$_{1-x}$La$_x$MnO$_3$ ($x\leq 0.10$)
were prepared by solid-state reaction as described
elsewhere.\cite{NeumeierCohn} Powder x-ray diffraction revealed
no secondary phases and iodometric titration, to measure the
average Mn valance, indicated an oxygen content
within the range 3.00$\pm$0.01 for all specimens.  The magnetization\cite{NeumeierCohn}
and thermal conductivity\cite{CohnNeumeier} of similar specimens have described previously.
A single crystal was grown from two polycrystalline rods of CaMnO$_3$
that were melted in an dual-mirror optical image furnace; one
acted as a seed, the other was the feed rod.  Both rods were
rotated in opposing directions at 50~rpm during the growth and the
rods moved downward at 5~mm/h through the hot zone.  The
heating power was 560~W and the growth was conducted in 2~atm. oxygen.
Small single crystals a few mm in length were cut
from the recrystallized rod.  Resistivity, Hall, and thermopower (TEP) measurements
were performed on all polycrystalline specimens; resistivity and TEP only for the crystal.
The TEP was measured using a steady-state technique employing a fine-wire
chromel-constantan thermocouple and gold leads.  Dc Hall and resistivity
measurements were performed in separate experiments on 6-probe
Hall-bar specimens of approximate dimensions $1\times 3\times 0.15\ {\rm mm}^3$
in a 9T magnet; both current and field reversal were employed in the Hall measurements.
%
\section{\label{sec:Results} RESULTS}

\begin{figure}
\vglue -.15in
\includegraphics[width = 3.4in, clip]{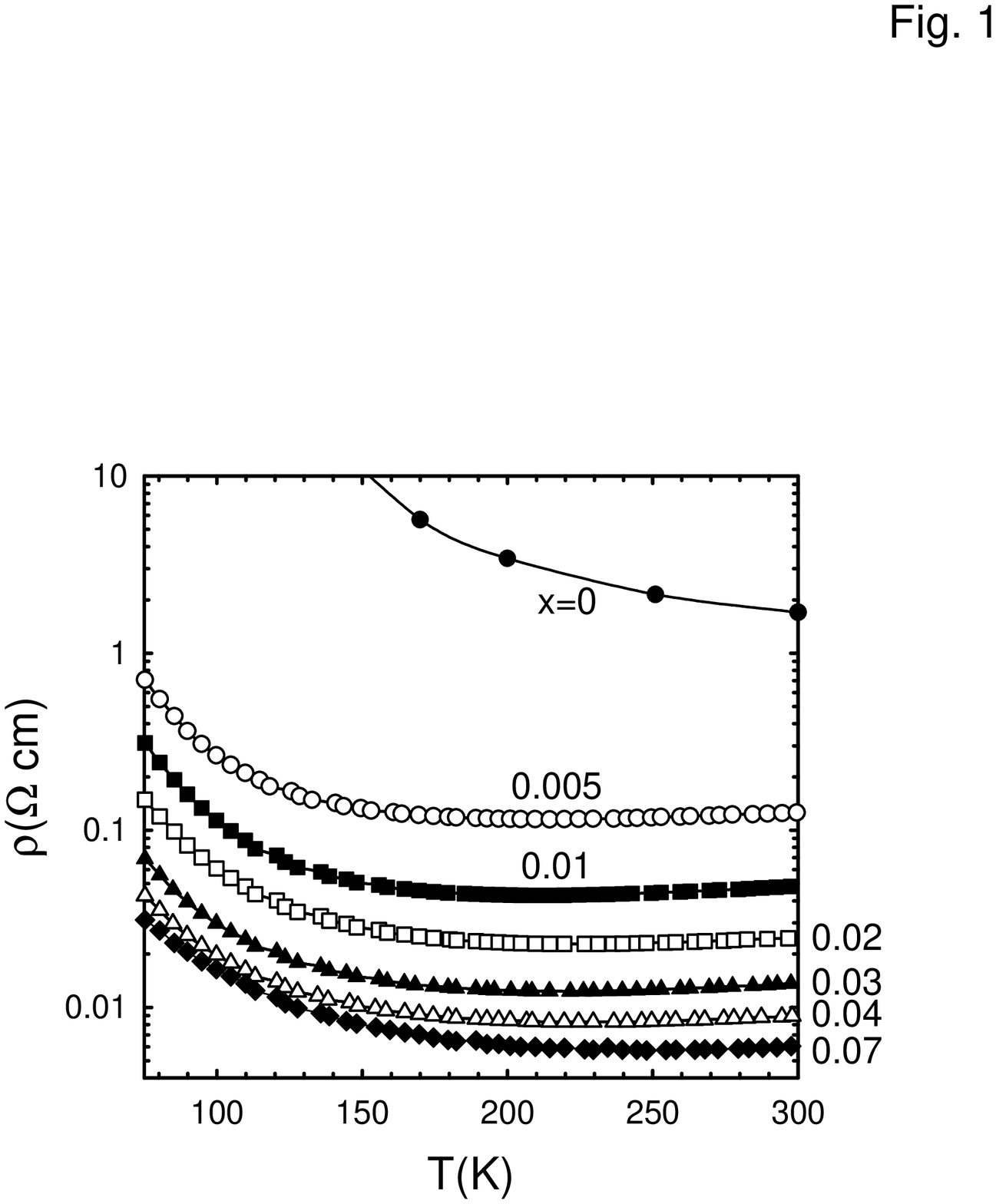}%
\vskip -.1in
\caption{\label{resistivity} Resistivity versus temperature for
Ca$_{1-x}$La$_x$MnO$_3$ in the paramagnetic phase. Data for $x=0.10$ are omitted for
clarity.}
\end{figure}

Figure~\ref{resistivity} shows the electrical resistivity versus temperature for
all of the compounds.  The magnitude of $\rho$ decreases systematically with doping.
Specimens with $0.005\leq x\leq 0.04$ exhibit a positive temperature coefficient of resistivity
for $T>200$~K.  Similar qualitative features have been reported for electron doping with other
lanthanides.\cite{Maignan,OtherRaveau,YCa}

\begin{figure}
\vglue -.18in
\includegraphics[width = 3in, clip]{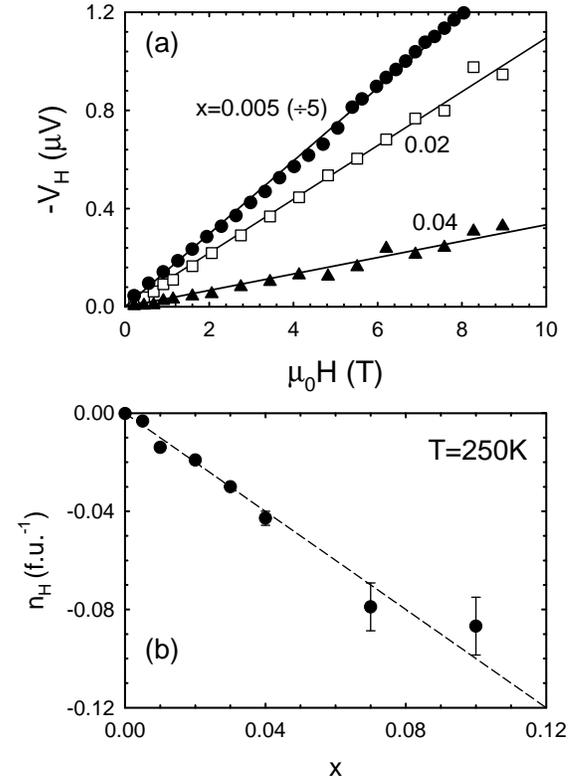}%
\vglue -.1in
\caption{\label{HallVoltage} a) Hall voltage versus magnetic field, and
(b) Hall number versus $x$ at $T=250$~K.}
\end{figure}

The Hall voltage, shown at $T=250$~K for several compositions in
Fig.~\ref{HallVoltage}~(a), was negative (electron-like) and
linear in field for all specimens and temperatures in the PM
phase.  The Hall coefficient was computed as,
$R_H=[dV_H/d(\mu_0H)](t/I)$, where $t$ is the specimen dimension
along the field and $I$ the current. Figure~\ref{HallVoltage}~(b)
shows the Hall number, $n_H=V_{f.u.}/R_H|e|$, as a function of
doping at $T=250$~K. $V_{f.u.}$ was taken as $(205/4)~{\rm \AA}^3$
for all $x$ since variations in the cell volume\cite{LingGranado}
over this range of doping constitute variations in $n_H$ smaller
than the measurement accuracy.  It is seen that $n_H=-x$ is well
obeyed, confirming that La doping adds approximately $x$ electrons
per formula unit.  A value of $n_H=0.01$ corresponds to a carrier
density of $\simeq 2.0\times 10^{20}$~cm$^{-3}$.

\begin{figure}
\vskip -.05in
\includegraphics[width = 3.2in, clip]{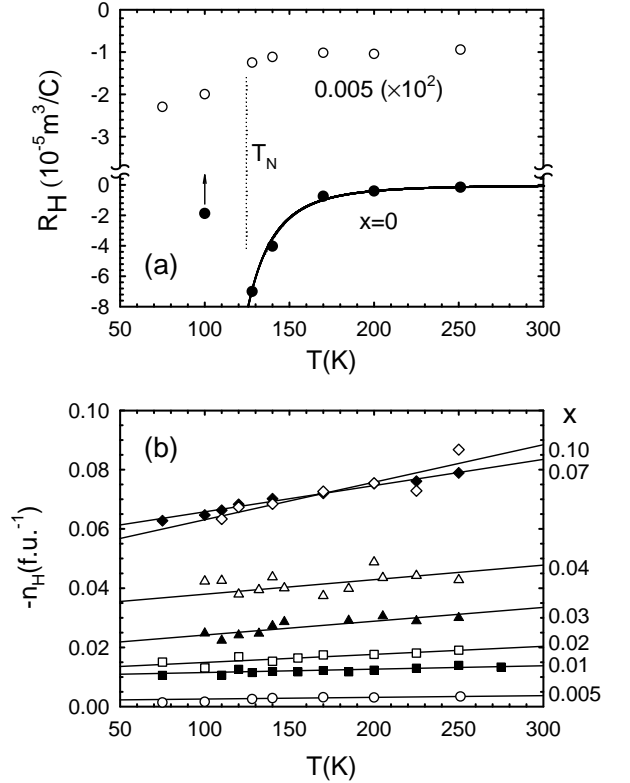}%
\vskip -.2in
\caption{\label{Hallnumber} (a) $R_H(T)$ for $x=0,\ 0.005$.  The
solid curve is an exponential fit (see text).  (b) $n_H(T)$ for
$x\geq 0.005$ with linear-$T$ fits (solid lines).}
\end{figure}
$R_H(T)$ has an exponential dependence for $x=0$ in the
paramagnetic phase with activation energy, $E_H/k_B\sim 1000$~K
[Fig.~\ref{Hallnumber}~(a); the solid curve corresponds to
$n_H=-1.25\times 10^{-2}\exp(-1010/T)$]. Below $T_N\approx 125$~K,
$R_H$ turns sharply toward a positive value at 75~K corresponding
to $n_H\simeq 4\times10^{-8}$ (this was the lowest $T$ for which
$R_H$ for $x=0$ could be reliably measured). This indicates
partial compensation by a small density of holes. A small oxygen
vacancy concentration is a likely source of electrons in
CaMnO$_3$, but a distribution of donors and acceptors is common in
oxides. A smaller concentration of acceptors in the present
compounds is expected to arise from several ppm levels of
impurities (e.g., Al, Zn) in the starting chemicals. A smaller
feature at $T_N$ is observed for the x=0.005 specimen: an increase
in $|R_H|$ with no sign change [Fig.~\ref{Hallnumber}~(a)]. The
$x=0.005$ specimen is degenerately doped with electrons so the
behavior of $R_H$ implies a decrease of the mobile electron
concentration at $T<T_N$. The Hall coefficient in the presence of
both mobile electrons ($n$) and holes ($p$) is expressed as,
$R_H=(1/e)(n\mu_n^2-p\mu_p^2)/(n\mu_n+p\mu_p)^2$, where $\mu_n$
and $\mu_p$ are the electron and hole mobilities, respectively.
Evidently as $n$ is suppressed below $T_N$, the hole-like
contribution predominates in the $x=0$ specimen.  No measurable
effects are detected upon crossing $T_N$ in $n_H(T)$ for the other
La-doped specimens [Fig.~\ref{Hallnumber}~(b)]. Their weak $T$
dependencies for $n_H$ are typical of degenerately-doped
semiconductors, and obey the empirical relation, $n_H=A+BT$ [solid
lines, Fig.~\ref{Hallnumber}~(b)].
\begin{figure}
\vglue -.15in
\includegraphics[width = 3.2in, clip]{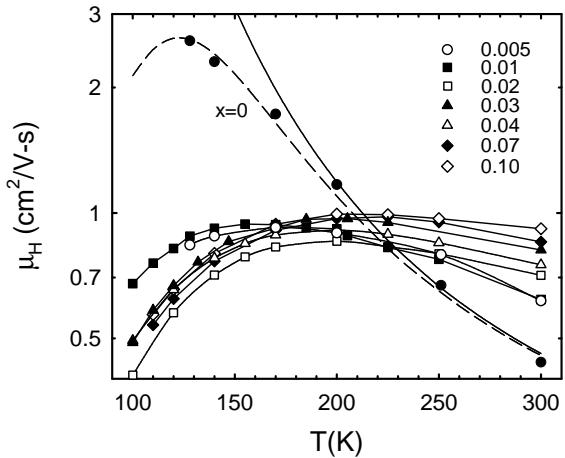}%
\caption{\label{HallMobility} $\mu_H(T)$ for all compositions in
semilog scaling. The solid curve for $x=0$ represents Feynman
polaron theory with phonon energy $\Theta=700$~K and
$m^{\ast}=4.3m_0$ (see text). The dashed curve is a fit using the
same polaron parameters with an additional term for impurity
scattering (see text). The curves for doped specimens are guides
to the eye.}
\end{figure}
\begin{figure}
\includegraphics[width = 3.4in, clip]{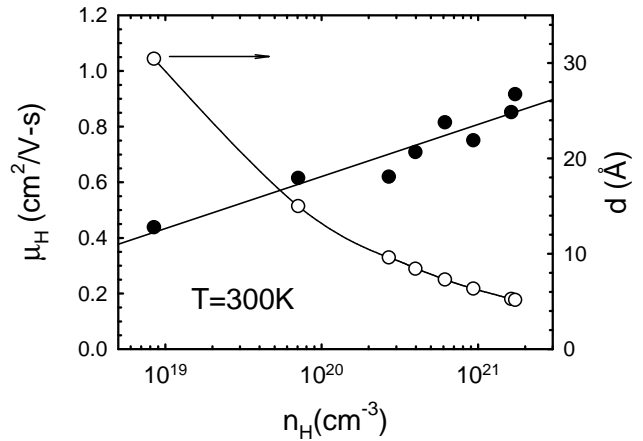}%
\caption{\label{MuvsN} $\mu_H$ (solid circles) and mean dopant spacing (open circles, right ordinate)
{\it vs} Hall carrier density at room temperature.  Curves are guides to the eye.}
\end{figure}

The Hall mobilities, $\mu_H\equiv R_H/\rho$ for all specimens are plotted
in Fig.~\ref{HallMobility}, where the fits from Fig.~\ref{Hallnumber}
have been employed to produce smoothed results.  In general, $\mu_H$ determined from polycrystalline specimens
should be viewed as a lower bound on intrinsic behavior, given that grain-boundary scattering tends to increase
$\rho$, but has little effect on $R_H$ even for highly anisotropic materials.\cite{HallTheory}  However our
single crystal of CaMnO$_3$, with a carrier density estimated
from the TEP (see below) to be between that of the $x=0$ and $x=0.005$ polycrystal specimens,
had a room-temperature resistivity of 2.7~$\Omega$~cm, somewhat larger than that of the $x=0$ polycrystal
and implying a lower value for $\mu_H$.  Thus
grain-boundary scattering appears to be relatively insignificant in determining the behavior of $\mu_H$.

$\mu_H(T)$ increases strongly with decreasing temperature for $x=0$ from a room-temperature value of
$\mu_H\simeq 0.5$~cm$^2$/V-s (Fig.~\ref{HallMobility}).  The doped specimen
mobilities increase more weakly with decreasing temperature, reach maxima in the range 150-200K,
and decrease at lower temperatures.  The qualitative features of the data are typical of polar semiconductors,
with an intrinsic regime associated with phonon scattering at high $T$ and an extrinsic region controlled
by charged impurity scattering (particularly for the doped specimens) at low-$T$.  The weak, systematic
increase in $\mu_H$ with doping at high $T$ (Fig.~\ref{MuvsN}) suggests another mechanism,
discussed further in the next section.  The overall magnitude,
$\mu_H\sim 1$~cm$^2$/V-s at 200~K, is nearly two orders of
magnitude larger than that of hole-doped FM compositions\cite{Jaime} and the small-polaron hopping
mobility for the latter [$\mu_H\propto \exp(-E/k_BT)$], contrasts
with the behavior observed here.

\begin{figure}
\vglue -.12in
\includegraphics[width = 3.2in, clip]{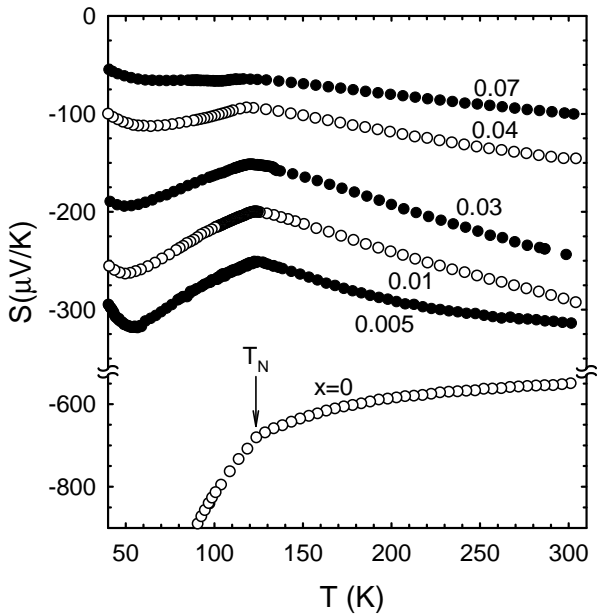}%
\caption{\label{TEP} Thermopower versus temperature. Data for $x=0.02$ and 0.10
are omitted for clarity.
}
\end{figure}

%
Thermoelectric power data are shown in Fig.~\ref{TEP}.  The TEP for $x=0$ is $\sim -550~\mu$V/K at room temperature
and grows larger with decreasing $T$, consistent with the presence of an activation energy as indicated by the
behavior of $\rho(T)$ and $n_H(T)$.  With
increasing $x$ the TEP is reduced in magnitude and has a $T$ dependence typical of degenerately-doped
semiconductors.\cite{Slack}
The overall features are similar to observations on Sm- and Pr-doped CMO.\cite{Maignan,OtherRaveau}
The TEP behavior for the doped specimens contrasts with the thermally activated ($S\propto 1/T$) behavior
that typifies the PM-phase TEP of hole-doped, FM compounds.\cite{Jaime,CohnReview,JaimeTEP}

The TEP magnitude for all specimens increases abruptly at $T<T_N$.  This is attributed to
a reduction in electron transfer (via double-exchange) as the majority of Mn core spins take on an AF arrangement.
The corresponding reduction in the mobile electron concentration enhances the TEP magnitude, in agreement
with our observations above regarding the Hall coefficient.  We also note that small features at $T_N$, consistent
with a decrease in carrier concentration and/or mobility, are evident in the temperature derivatives of
the $\rho(T)$ curves.

For all of the doped specimens, the TEP tends toward zero as $T\to 0$, indicating a finite density of
states at the Fermi level.  This is consistent with prior low-$T$ resistivity\cite{NeumeierCohn} and
ac impedance measurements\cite{CohnCSMOEps} which suggest that the Fermi level lies in a
disorder-broadened impurity ``band'' close in energy to the band edge.

\section{\label{sec:A&D} ANALYSIS AND DISCUSSION}

A strong interaction between carriers and optical phonons typifies perovskite oxides.
As noted above, the magnitude and temperature dependence of both $\mu_H$ and the TEP
argue against the use of small (Holstein) polaron theory which has been successful in describing hole-doped manganites.
Nevertheless, a large static dielectric constant\cite{CohnCSMOEps}
$\varepsilon_0\simeq 40$, determined for similarly prepared CMO, makes polaron formation likely
in electron-doped compounds given that the optical-frequency dielectric
constant for manganites (and oxides generally)\cite{EpsInfty}
is substantially smaller, $\varepsilon_{\infty}\simeq 5$.  These observations motivate a Fr\"ohlich
(continuum or large) polaron description for the charge carriers in this system.  Before analyzing the data, it is
useful to estimate the dimensionless polaron coupling constant,\cite{Seeger}
\begin{equation*}
\alpha=397.4\left[(m^{\ast}/m_0)\over\Theta\right]^{1/2}(\varepsilon_{\infty}^{-1}-\varepsilon_{0}^{-1}),
\end{equation*}
\noindent where $m^{\ast}$ is the band mass (without polaron
enhancement), $m_0$ the free-electron mass, $\Theta=\hbar\omega_0/k_B$, and $\omega_0$ the
longitudinal optic (LO) phonon frequency.  Taking $\Theta\sim 700$~K as
an average LO phonon energy,\cite{RamanIR} and $m^{\ast}=m_0$ implies
$\alpha\sim 2.6$.  The donor binding energy and dielectric constant of CMO suggest\cite{CohnCSMOEps}
$m^{\ast}\sim 4m_0$ and thus a proper treatment of the transport
properties requires a theory suitable for intermediate coupling
($2\leq \alpha\leq 6$).

The most reliable theory for large polaron mobility at intermediate
coupling and intermediate temperatures ($T\lesssim \Theta$) is
that of Feynman {\it et al.}\cite{Feynman,Biaggio} The polaron
mobility is given (in cm$^2$/V~s) as   ,
\begin{equation*}
\mu_p={7.14\times10^4\over \alpha\Theta}\left(m_0\over
m^{\ast}\right){\sinh(z/2)\over (z/2)^{5/2}}{w^3\over v^3}{1\over
K(v,w,z)}.
\end{equation*}
\noindent where $z=\Theta/T$.  The integral $K(v,w,z)$ and the
procedure for determining the Feynman variational parameters $v$
and $w$ at each temperature for a given value of $m^{\ast}/m_0$
are described in Ref.~\onlinecite{Biaggio}. The mobility data at
the highest temperatures for $x=0$ were fitted using
$\Theta=700$~K and $m^{\ast}/m_0=4.3$ ($\alpha\simeq 5.4$)[solid
curve in Fig.~\ref{HallMobility}].  The discrepancy between experiment and theory below 175~K is
plausibly attributed to the growing role of impurity scattering with
decreasing $T$.  To fit the entire $T$ range, $\mu_p$ was combined
with the Brooks-Herring mobility\cite{BHMobility} for charged
impurity scattering (in cm$^2$/V~s),
\begin{eqnarray*}
\mu_{BH}=&&{3.68\times10^{20}{\rm cm}^{-3}\over N_I}{1\over Z^2}
\left(\varepsilon_0\over 16\right)^2\left(T\over 100
K\right)^{3/2}\\
&&\times\left(m_0\over m^{\ast}\right)^{1/2}f(\beta)
\end{eqnarray*}
\begin{equation*}
f(\beta)=[\ln(1+\beta^2)-0.434\beta^2/(1+\beta^2)]^{-1}
\end{equation*}
\begin{equation*}
\beta=\left(\varepsilon_0\over 16\right)^{1/2}\left(T\over 100
K\right)\left(m^{\ast}\over m_0\right)^{1/2}
\left(2.08\times10^{18}{\rm cm}^{-3}\over n\right)^{1/2}
\end{equation*}
The dashed curve through the $x=0$ mobility (Fig.~\ref{HallMobility}) represents
$\mu=(\mu_p^{-1}+\mu_{BH}^{-1})^{-1}$, using $n=n_H(T)$, $Z=2$ (for oxygen vacancies),
$\varepsilon_0=40$, and $m^{\ast}/m_0=4.3$.  The impurity concentration (the remaining free parameter) required to
fit the data was $N_I=3\times 10^{19}$~cm$^{-3}$, roughly four times the $T=300$~K Hall carrier concentration.
A similar discrepancy was observed\cite{Chen95} for the impurity term describing the
mobility of La$_2$CuO$_{4+y}$.  The disagreement is reasonable considering that only the charge difference
associated with the oxygen vacancy, but not the lattice distortion, is considered in the Brooks-Herring theory.
Long-ranged, correlated disorder is an important characteristic\cite{Burgy} of the perovskite oxides due to the
corner-sharing metal-oxygen polyhedra.

Let us briefly discuss the polaron mass and size dictated by this analysis.
The Feynman polaron mass at room temperature is $m_p=(v/w)^2m_0\simeq 36m_0$.
The Feynman path integral theory gives a larger effective mass than other
approaches to the problem and a reasonable lower bound on the mass is given by
the perturbation theory result, shown to be a good approximation
even for intermediate coupling,\cite{AlexandrovMott}
$m_p=m^{\ast}(1+\alpha/6)\simeq 8.4m_0$.  The Feynman polaron radius is given
as,\cite{Schultz} $R_p=(3\hbar/2\mu v)^{1/2}\simeq 3{\rm\AA}$ [$\mu=m^{\ast}(v^2-w^2)/v^2$],
comparable to a lattice spacing.  Thus the continuum approximation is near its limit
of validity.  This value of $R_p$ is consistent with the 7-site FM
polaron predicted for the magnetically-ordered ($T=0$) state of CaMnO$_3$
from theoretical studies\cite{ChenAllen,Meskine} incorporating both
lattice and spin interactions.  Magnetic contributions to PM-phase polaron formation
are presumably less significant than that of the lattice, though it has been proposed that
magnetic fluctuations increase the polaron binding energy, yielding low-mobility magneto-elastic
polarons for lower electron-phonon coupling strengths.\cite{Nagaev}  The latter effects could
possibly contribute to the downturn in the mobility modeled above by charged impurity scattering.

The simple sum of scattering mechanisms describing the $\mu_H(T)$ for $x=0$ is inadequate for the doped
specimens, given the systematic increase in $\mu_H$ with doping at high $T$ (Fig.~\ref{MuvsN}).  The enhanced
mobility might signify the effects of overlapping polaronic lattice distortions.  If so, the mean
distance between dopants, $d=(3/4\pi n)^{1/3}$ ($n$ is the carrier density) plotted in
Fig.~\ref{MuvsN} (open circles), suggests
that the distortion field extends over a few lattice spacings.  Low-$T$ magnetization\cite{NeumeierCohn} and
neutron scattering\cite{LingGranado} studies suggest a crossover near $x=0.02$ from isolated
to interacting FM polarons.

The TEP offers an alternative, though less reliable, means for assessing
the effective mass.  We are not aware of a theory for the TEP valid for
intermediate coupling, but the weak-coupling (perturbative) theory of
Howarth and Sondhemier (HS)\cite{HowarthSondheimer} was found to
provide a good description of the TEP. It is known that weak
coupling theories tend to overestimate the band effective mass.  A
heuristic approach, wherein the band mass within weak-coupling
theories is interpreted as the polaron-enhanced mass, has been
proposed to broaden their range of applicability.\cite{GarciaMoliner}
We restrict our analysis to the
doping dependence of the TEP at room temperature where optical
phonon scattering is predominant. Though the HS theory reproduces
the $T$ dependence of the TEP well, such an analysis is less
meaningful given that the theory does not incorporate impurity
scattering (of growing importance for the present compounds at low $T$) or a temperature
dependent polaron mass (implicit in Feynman theory).

\begin{figure}
\includegraphics[width = 3.2in, clip]{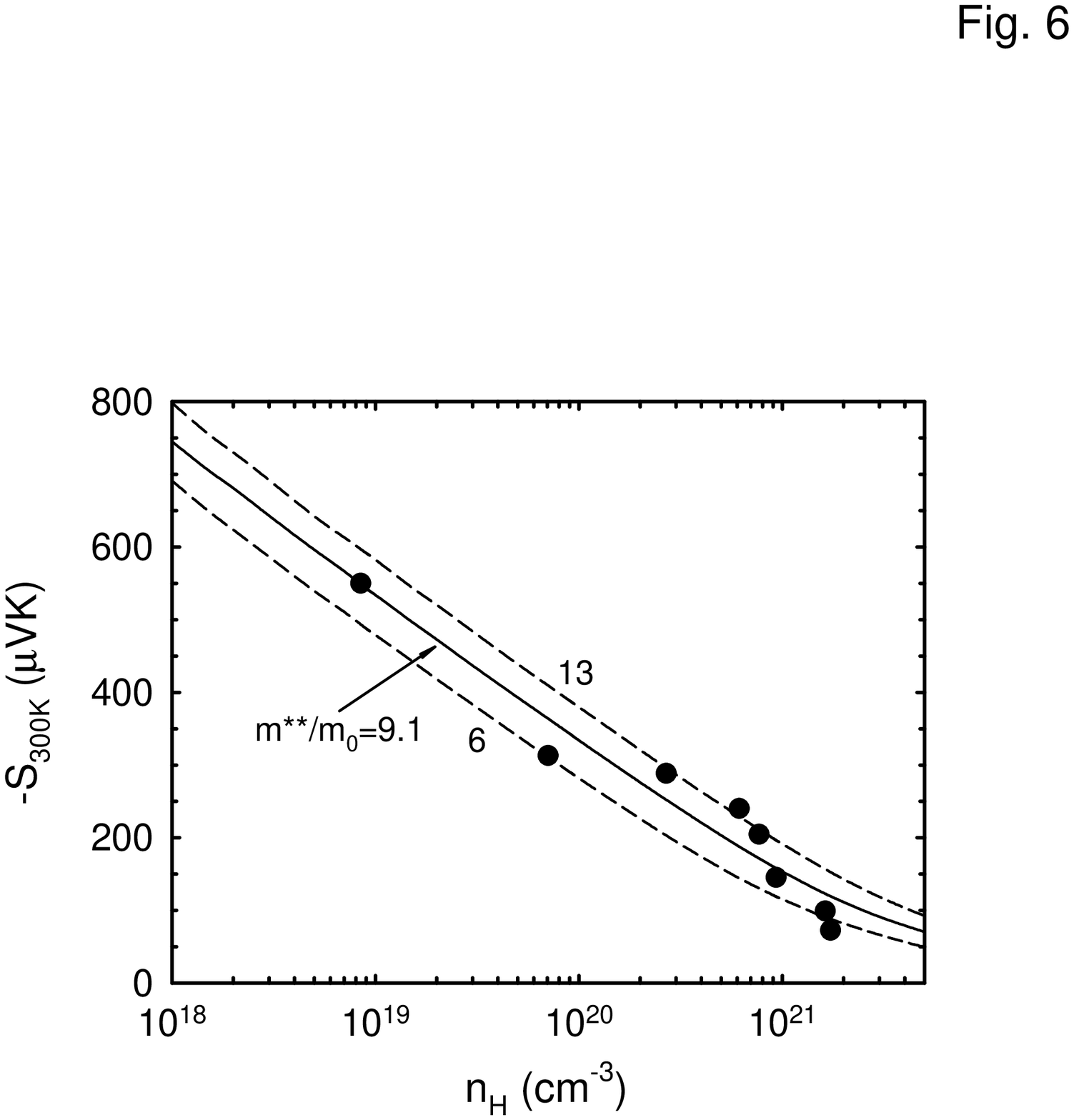}%
\caption{\label{TEPvsN} Room-temperature thermopower {\it vs.} Hall carrier density.  The solid and dashed curves
were computed using the theory of Ref.~\onlinecite{HowarthSondheimer}, with effective masses
indicated (see text).}
\end{figure}

The HS theory gives the TEP as a function of the phonon energy
($\Theta$) and reduced chemical potential.\cite{HSNote} With
$\Theta=700$~K and the chemical potential determined
self-consistently from the Hall carrier density (assuming a
parabolic band), the TEP can be computed with the effective mass
(designated here as $m^{\ast\ast}$) as the only adjustable
parameter.  Figure~\ref{TEPvsN} shows the room temperature TEP plotted versus
Hall carrier density for all specimens.  The solid curve represents the best fit to the
HS theory with $m^{\ast\ast}=9.1m_0$.
Variations in the polaron mass with doping might possibly explain the discrepancy
between the calculated and measured TEP magnitudes (corresponding to variations in
$m^{\ast\ast}/m_0$ from 6 to 13, indicated by dashed curves), but these differences
may simply reflect a sensitivity of the TEP to extrinsic features, e.g. small variations
in the oxygen vacancy concentration (below the uncertainty of 0.01 per f.u. from titration),
that cause it to deviate from being a monotonic function of the carrier density as reflected in
the Hall coefficient.  Thus the polaron masses determined from $\mu_H$ and the TEP
are in reasonable accord.

It is useful to place in context the intermediate-coupling, Fr\"ohlich polaron picture for the paramagnetic phase of
electron-doped manganites favored by the present analysis.  Recently, polaron parameters were inferred from
pressure-dependent resistivity measurements\cite{Garbarino} on Ca$_{1-x}$Y$_x$MnO$_3$ ($0.05\leq x\leq 0.15$)
having $\rho(T)$ very similar to that of the most heavily doped specimens in Fig.~\ref{resistivity}.
These authors interpreted the minimum in $\rho(T)$ (near $T=200$~K in Fig.~\ref{resistivity})
as a manifestation of the crossover from
the intermediate-$T$ small-polaron regime, characterized by thermally activated hopping, to the high-$T$
regime where polarons are thermally dissociated and phonon scattering of residual electrons yields a positive
temperature coefficient of $\rho$.\cite{Appel,Fratini}  Analyzing the pressure dependence of the
activation energy from $\rho(T)$ over the narrow $T$ interval between $T_N$ and the minimum, they inferred
a coupling strength, $\alpha\simeq 1.5$.  The strongest argument against this interpretation is that the minimum
in $\rho(T)$ arising in small polaron theory occurs at several times the optical phonon energy ($\Theta$),
well above the temperature range investigated.  In addition, as noted above, polarons with
activated $\rho(T)$ are expected to exhibit an activated TEP ($\propto 1/T$), inconsistent with the present
observations and those of others.\cite{Maignan,OtherRaveau}

The electron-doped manganite compounds are evidently near a large- to small-polaron crossover.  A comparable
situation arises in TiO$_2$ (rutile), for which $\mu_H(T)$ and $\rho(T)$ are similar to
data for the present materials and both small\cite{AlexandrovMott,Bogomolov,Yagi} and
large\cite{Baumard,Hendry} polaron descriptions have been invoked.
Recent measurements of the Hall mobility by an all-optical technique\cite{Hendry} support a
continuum picture for that system, like that described here.  The appropriate theoretical
framework for describing such materials remains an active area of
investigation,\cite{Fratini,Nagaev,AlexandrovKorn,Perroni} and the electron-doped compounds appear
to be model systems for studying polaron physics.

In summary, the paramagnetic-phase Hall mobility and thermopower of
Ca$_{1-x}$La$_x$MnO$_3$ ($0\leq x\leq 0.10$) and other
lightly electron-doped manganites by implication, are consistent with large (continuum) polaron theory
in the intermediate coupling regime.  This behavior is distinguished from the small-polaron scenario that
has been successful in describing the paramagnetic phase of hole-doped FM manganites.

\section{\label{sec:Ack} ACKNOWLEDGMENTS}

The work at the Univ. of Miami was supported by NSF Grant No. DMR-0072276, and at
Montana State Univ. by Grant No. DMR-0301166.


\begin{references}
\bibitem{Dagotto} E. Dagotto, T. Hotta, and A. Moreo, Phys. Rep.
{\bf 344}, 1 (2001).
\bibitem{Jaime} M. Jaime, H. T. Hardner, M. B. Salamon, M. Rubenstein, P. Dorsey, and D. Emin,
Phys. Rev. Lett. {\bf 78}, 951 (1997).
\bibitem{CohnReview} J. L. Cohn, J. Supercond:Incorp. Nov. Magn., {\bf 13}, 291 (2000).
\bibitem{SalamonJaime} M. B. Salamon and M. Jaime, Rev. Mod. Phys. {\bf 73}, 583 (2001).
\bibitem{PolaronEnergy} J. M. De Teresa, K. D\"orr, K. H. M\"uller, L. Schultz, and R. I. Chakalova,
Phys. Rev. B {\bf 58}, R5928 (1998).

\bibitem{Maignan} A. Maignan, C. Martin, F. Damay, B. Raveau,
and J. Hejtm\'anek, Phys. Rev. B {\bf 58}, 2758 (1998).
\bibitem{NeumeierCohn} J. J. Neumeier and J. L. Cohn, Phys. Rev. B {\bf 61},
 14319 (2000).
\bibitem{OtherRaveau} C. Martin, A. Maignan,
M. Herviieu, B. Raveau, Z. Jirák, M. Savosta, A. Kurbakov, V. Trounov,
G. André, and F. Bourée, Phys. Rev. B {\bf 62}, 6442 (2000); M. M. Savosta,
P. Novák, M. Marysko, Z. Jirák, J. Hejtmánek, J. Englich, J. Kohout, C. Martin,
and B. Raveau, {\it ibid.}, 9532 (2000)
\bibitem{YCa} H. Aliaga, M. T. Causa, M. Tovar, and B. Alascio, Physica B {\bf 320}, 75 (2002);
H. Aliaga, M. T. Causa, B. Alascio, H. Salva, M. Tovar, D. Vega, G. Polla, G. Leyva, and P. Konig,
J. Magn. Magn. Mater. {\bf 226-230}, 791 (2001).
\bibitem{CohnNeumeier} J. L. Cohn and J. J. Neumeier, Phys. Rev.
B {\bf 66}, 100404 (2002).
\bibitem{LingGranado} C. D. Ling, E. Granado, J. J. Neumeier, J. W. Lynn, and D. N.
Argyriou, Phys. Rev. B {\bf 68}, 134439 (2003); E. Granado, C.
D. Ling, J. J. Neumeier, J. W. Lynn, and D. N. Argyriou, {\it
ibid.}, 134440 (2003).

\bibitem{HallTheory} Ting-Kang Xia and D. Stroud, Phys. Rev. B {\bf 37}, 118 (1988).

\bibitem{Slack} G. A. Slack and M. H. Hussain, J. Appl. Phys. {\bf 70}, 2694 (1991).

\bibitem{JaimeTEP} M. Jaime, M. B. Salamon, M. Rubenstein, R. E. Treece, J. S. Horwitz, and D. B. Chrisey,
Phys. Rev. B {\bf 54}, 11914 (1996).

\bibitem{CohnCSMOEps} J. L. Cohn, M. Peterca, and J. J. Neumeier, Phys. Rev. B {\bf 70}, 214433 (2004).

\bibitem{EpsInfty} A. S. Alexandrov and A. M. Bratkovsky, J. Phys.: Condens. Matter
{\bf 11}, L531 (1999).

\bibitem{Seeger} See, e.g., K. Seeger, {\it Semiconductor Physics}, (Springer-Verlag, Berlin, 1997), p. 209.

\bibitem{RamanIR} L. Kebin, L. Xijun, Z. Kaigui, Z. Jingsheng, and Z. Yuheng, J. Appl. Phys. {\bf 81}, 6943 (1997);
M. V. Abrashev, J. B\"ackstrom, L. B\"orjesson, V. N. Popov, R. A. Chakalov, N. Kolev, R.-L. Meng,
and M. N. Iliev, Phys. Rev. B {\bf 65}, 184301 (2002); L. Mart\'in-Carr\'on, A. de Andr\'es, M. J. Mart\'inez-Lope,
M. T. Casais, and J. A. Alonso, Phys. Rev. B {\bf 66}, 174303 (2002); N. N. Loshkareva, L. V. Nomerovannaya, E. V. Mostovshchikova, A. A. Makhnev, Yu. P. Sukhorukov, N. I. Solin,
Y. I. Arbuzova, S. V. Naumov, N. V. Kostromitina, A. M. Balbashov, and L. N. Rybina, Phys. Rev. B {\bf 70}, 224406 (2004).

\bibitem{Feynman} R. P. Feynman, Phys. Rev. {\bf 97}, 660 (1955); R. P. Feynman, R. W. Hellwarth, C. K. Iddings,
and P. M. Platzman, {\it ibid.} {\bf 127}, 1004 (1962).

\bibitem{Biaggio} R. W. Hellwarth and I. Baggio, Phys. Rev. B {\bf
60}, 299 (1999).

\bibitem{BHMobility} K. Seeger, {\it Semiconductor Physics,} 6th Ed. (Springer-Verlag, New York, 1997), Ch. 6;
D. Chattopadhyay and H. J. Queisser, Rev. Mod. Phys. {\bf 53}, 745 (1981).

\bibitem{Chen95} C. Y. Chen, E. C. Branlund, ChinSung Bae, K. Yang,
M. A. Kastner, A. Cassanho, and R. J. Birgeneau , Phys. Rev. B {\bf 51}, 3671 (1995).

\bibitem{Burgy} J. Burgy, A. Moreo, and E. Dagotto, Phys. Rev. Lett. {\bf 92}, 097202 (2004).

\bibitem{AlexandrovMott} A. S. Alexandrov and N. Mott, {\it Polarons and Bipolarons,} (World Scientific, Singapore, 1995).

\bibitem{ChenAllen} Y.-R. Chen and P. B. Allen, Phys. Rev. B {\bf 64}, 064401 (2001).

\bibitem{Meskine} H. Meskine, T. Saha-Dasgupta, and S. Satpathy, Phys. Rev. Lett. {\bf 92}, 056401 (2004).

\bibitem{Nagaev} E. L. Nagaev, Phys. Rev. B {\bf 60}, R6984 (2001).

\bibitem{Schultz} T. D. Schultz, Phys. Rev. {\bf 116}, 526 (1959).

\bibitem{HowarthSondheimer} D. J. Howarth and E. H. Sondheimer, Proc. Roy. Soc. London, {\bf A 219}, 53 (1953).

\bibitem{GarciaMoliner} F. Garcia-Moliner, Phys. Rev. {\bf 130}, 2290 (1963).

\bibitem{HSNote} Equations 70-73 from Ref.~\onlinecite{HowarthSondheimer} for the case of
arbitrary degeneracy were employed.

\bibitem{Garbarino} G. Garbarino, C. Acha, D. Vega, G. Leyva, G. Polla, C. Martin, A. Maignan, and B. Raveau,
Phys. Rev. B {\bf 70}, 014414 (2004).

\bibitem{Appel} J. Appel, in {\it Solid State Physics}, edited by F. Seitz, D. Turnbull, and H. Ehrenreich
(Academic Press, New York, 1968), Vol. 21, pp. 193-391.

\bibitem{Fratini} S. Fratini and S. Ciuchi, Phys. Rev. Lett. {\bf 91},
256403 (2003).

\bibitem{Bogomolov} V. N. Bogomolov, E. K. Kudinov, and Yu A. Firsov, Fiz. Tverd. Tela {\bf 9}, 3175 91967).

\bibitem{Baumard} J. F. Baumard and F. Gervais, Phys. Rev. B {\bf 15}, 2316 (1977).

\bibitem{Yagi} E. Yagi, R. R. Hasiguti, and M. Aono, Phys. Rev. B {\bf 54}, 7945 (1996).

\bibitem{Hendry} E. Hendry, F. Wang, J. Shan, T. F. Heinz, and M. Bonn, Phys. Rev. B {\bf 69} 081101(R) (2004).

\bibitem{AlexandrovKorn} A. S. Alexandraov and P. E. Kornilovitch, Phys. Rev. Lett. {\bf 82}, 807 (1999).

\bibitem{Perroni} C. A. Perroni, G. Iadonisi, and V. K. Mukhomorov, cond-mat/0411653.

\end{references}
\end{document}